# Heuristic to reduce the complexity of complete bipartite graphs to accelerate the search for maximum weighted matchings with small error

Daniel Etzold
etzold@cs.tum.edu

August 9, 2018

**Abstract**

A maximum weighted matching for bipartite graphs $G = (A \cup B, E)$ can be found by using the algorithm of Edmonds and Karp [1] with a Fibonacci Heap and a modified Dijkstra in $\mathcal{O}(nm + n^2 \log n)$ time where $n$ is the number of nodes and $m$ the number of edges. For the case that $|A| = |B|$ the number of edges is $n^2$ and therefore the complexity is $\mathcal{O}(n^3)$.

In this paper we want to present a simple heuristic method to reduce the number of edges of complete bipartite graphs $G = (A \cup B, E)$ with $|A| = |B|$ such that $m = n \log n$ and therefore the complexity of the maximum weighted matching algorithm will be $\mathcal{O}(n^2 \log n)$. The weights of all edges in $G$ must be uniformly distributed in $[0, 1]$.

# 1 Introduction

Computing the maximum or minimum weighted matching of bipartite graphs plays an important role in practice. Two applications are the image feature matching problem [2] or the assignment problem.

A well known algorithm is due to Edmonds and Karp and dates back to the year 1972. The two key ideas behind their network flow algorithm which converts the problem into a network flow problem is a) use the BFS to find augmenting paths and b) find a path that increases the flow at most.

All logarithms in this paper are to the base of 2 and all graphs $G = (A \cup B, E)$ have the following properties: each partition $A$ and $B$ of $G$ consists of $n$ nodes and the weights of the edges in $E$ are uniformly distributed over the interval [0,1]. Therefore, the number of edges is $m = n^2$. To reduce the complexity of the maximum weighted matching algorithm by Edmonds and Karp to $\mathcal{O}(n^2 \log n)$ we have to reduce the number of edges in $E$ to $n \log n$. Therefore we have to remove $n^2 - n \log n$ edges from $E$. The heuristic presented here will do this in $\mathcal{O}(n^2)$ time.

The paper is organized as follows: in section 2 we describe a simple approach to reduce $m$ to $n \log n$, in section 3 we try to improve the heuristic a bit and in section 4 we present empirical results for some graphs.

## 2 Heuristic

The heuristic is based on a randomized algorithm. This algorithm selects edges at random depending on their weight and builds a new graph $G' = (A \cup B, E')$ from $G$ where we expect $|E'| \approx n \log n$. Let $E = \{e_1, \ldots, e_{n^2}\}$. Then, we can compute the expected number of edges which will be selected by the algorithm to be in $E'$ as follows:

Let $X_i$ be a random variable for $1 \leq i \leq n^2$ indicating whether an edge $e \in E$ also is member of $E'$:

$$X_i := \begin{cases} 1, & e_i \in E' \\ 0, & e_i \notin E' \end{cases}$$

Furthermore, define $X := X_1 + X_2 + \ldots + X_{n^2}$. As we can see this is the number of edges in $E'$. The expected value of $X$ is computed as follows:

$$\mathbb{E}[X] = \sum_{i=1}^{n^2} \mathbb{E}[X_i] = \sum_{i=1}^{n^2} (1 \cdot \Pr[e \in E'] + 0 \cdot \Pr[e \notin E']) = \sum_{i=1}^{n^2} \Pr[e \in E']$$

where $\Pr[e \in E']$ is the probability that $e \in E'$.

A simple approach to compute the probability is to set $\Pr[e \in E'] = cw(e)$ where $w(e)$ is the weight of the edge $e$ and $c$ a constant. Then, we have to choose $c$ such that

$$\sum_{i=1}^{n^2} \Pr[e \in E'] = \sum_{i=1}^{n^2} cw(e_i) = n \log n$$

holds. This is done as follows:

$$\sum_{i=1}^{n^2} cw(e_i) = n \log n \quad \Rightarrow \quad c = \frac{n \log n}{\sum_{i=1}^{n^2} w(e_i)} \approx 2 \frac{\log n}{n}$$

The constant $c$ can be approximated by $2 \log n / n$ because the edges are uniformly distributed over the interval $[0, 1]$ and therefore, the value of the denominator is $\approx n^2/2$. Now, we can set the probability that an edge $e \in E$ is selected by the algorithm to

$$\Pr[e \in E'] = 2w(e) \frac{\log n}{n}.$$

Then, the expected number of edges in $E'$ is $n \log n$.

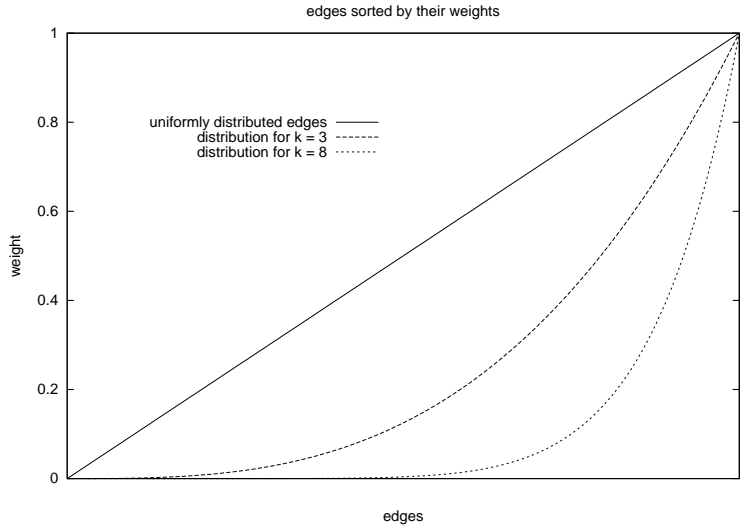

Figure 1: Graphs when sorting all edges of $E$ by their weights. For uniformly distributed weights we expect the upper graph. For weights modified by the heuristic described in section 2 we expect the lower graphs.

## 3 Modifying weights to improve results

In this section we want to modify the computation of the probability a bit to try to improve the heuristic of the previous section. The idea is not to use a constant factor to modify the weight of an edge but a function which computes a new weight $w'(e)$ for all $e \in E$ in the interval $[0, 1]$ depending on the weight of $e$. After applying this function to the weights of all edges of $E$ the new weights are not uniformly distributed anymore. Weights of edges near to one will be more probable than edges with small weights because we think that these edges are more important for a maximum weighted matching. The function is parameterized by a variable $k$.

We compute the new weight $w'(e)$ and the probability that the edge $e$ is in $E'$ as follows:

$$w'(e) = cw^k(e) \quad \Rightarrow \quad \Pr[e \in E'] = cw'(e) = cw^k(e)$$

As we can see for $k = 1$ this function is equal to the function presented in section 1 and the new weights are uniformly distributed.

The value for the constant $c$ can be computed analog to the method described in section 1.

$$\sum_{i=1}^{n^2} cw^k(e_i) = n \log n \quad \Rightarrow \quad c = \frac{n \log n}{\sum_{i=1}^{n^2} w^k(e_i)}$$

The value of $c$ can be approximated with the following idea. First, we sort all edges of $E$ by their new weight and put them into a diagram as shown in Figure 1. With the help of this diagram it is easy to see that the denominator of $c$ can be approximated by the area under the appropriate

| nodes | optimal | $k=5$ | $k=10$ | $k=20$ | $k=30$ | $k=40$ | $k=50$ |
|---|---|---|---|---|---|---|---|
| 1000 | 998.329 | 0.9691 | 0.9845 | 0.9916 | 0.9952 | 0.9967 | 0.9977 |
| 2000 | 1998.35 | 0.9720 | 0.9852 | 0.9920 | 0.9952 | 0.9965 | 0.9974 |
| 3000 | 2998.36 | 0.9732 | 0.9849 | 0.9924 | 0.9951 | 0.9964 | 0.9973 |
| 4000 | 3998.34 | 0.9746 | 0.9863 | 0.9939 | 0.9953 | 0.9965 | 0.9973 |
| 5000 | 4998.37 | 0.9753 | 0.9865 | 0.9930 | 0.9954 | 0.9966 | 0.9973 |
| 6000 | 5998.34 | 0.9757 | 0.9868 | 0.9931 | 0.9954 | 0.9966 | 0.9973 |

Figure 2: For six graphs the weights of optimal matchings and weights using the heuristic with different values for the parameter $k$ were computed. For weights which were computed by using the heuristic the tolerance to the optimal weight is given.

graph with the following integral:

$$\sum_{i=1}^{n^2} w^k(e_i) \approx \int_0^{n^2} \left(\frac{x}{n^2}\right)^k dx = \frac{1}{n^{2k}} \int_0^{n^2} x^k \, dx = \frac{1}{k+1} \cdot \frac{1}{n^{2k}} \left[x^{k+1}\right]_0^{n^2} =$$

$$\frac{1}{k+1} \cdot \frac{1}{n^{2k}} \cdot n^{2(k+1)} = \frac{1}{k+1} n^2$$

With that it follows

$$c = \frac{n \log n}{\sum_{i=1}^{n^2} w^k(e_i)} \approx \frac{n \log n}{\frac{1}{k+1} n^2} = (k+1) \frac{\log n}{n}$$

and

$$\Pr[e \in E'] \approx (k+1) \frac{\log n}{n} w^k(e)$$

As we can see the complexity of this heuristic is $\mathcal{O}(kn^2)$ because to each edge of $E$ we have to compute the new weight which is done by $k-1$ multiplications. Since we use fixed precision for all numbers we assume that each multiplication can be performed in $\mathcal{O}(1)$.

## 4 Empirical results

In this section we want to present some empirical results of our heuristic method for random graphs with 1000 up to 6000 nodes. For each graph we have computed the exact weight of a maximum weighted matching and the weight of a matching computed by our heuristic for several values for the parameter $k$. For the heuristic the weights are given relative to the exact weight. The results are shown in Figure 2.

The time required to find a maximum weighted matching is not shown in Figure 2 but for the graph with 6000 nodes and $k=50$ the network flow algorithm by Edmonds and Karp runs with the heuristic more than 22 times faster than without using the heuristic. Also, we recognized that smaller values for $k$ didn't boost the algorithm.